\documentclass[twocolumn,aps,prl,showpacs]{revtex4}

\usepackage{epsf}
\usepackage{bm}
\begin{document}

\title {`Magic Melters' Have Geometrical Origin.}

\author{Kavita Joshi, Sailaja Krishnamurty, and D. G. Kanhere}

\affiliation{%
Department of Physics and Center for Modeling and Simulation,
University of Pune, Ganeshkhind, Pune--411 007, India.}

\date{\today}


\begin{abstract}
Recent experimental reports bring out extreme size sensitivity 
in the heat capacities of Gallium and Aluminum clusters. In the present work 
we report results of our extensive {\it ab initio} molecular dynamical simulations
on Ga$_{30}$ and Ga$_{31}$, the pair which has shown rather dramatic size sensitivity.
We trace the origin of this size sensitive heat capacities to the relative order in
their respective ground state geometries. Such an effect of nature of
the ground state on the characteristics of heat capacities is also seen
in case of small Gallium and Sodium clusters indicating that  the observed 
size sensitivity  is a  generic feature of small clusters.

\end{abstract}

\maketitle

The finite temperature behavior of clusters has shown many interesting and 
intriguing properties~\cite{Na-expt,Jar-tin1,Ga-expt,Ga-JACS,Al-expt}.
Recently the calorimetric measurements reported by Jarrold and coworkers found that 
small clusters of Tin and Gallium  in the size range of 17-55 atoms have {\it higher than 
bulk} melting temperatures (Tm$_{[\rm bulk]}$)~\cite{Jar-tin1,Ga-expt}.
A striking experimental result from the same group showed extreme size sensitivity in 
the nature of the heat capacity for Ga clusters in the size range of 30-55 atoms~\cite{Ga-JACS}.
It turns out that addition of even one atom changes the heat capacity dramatically. 
For example, Ga$_{30}^{+}$ has a rather flat specific heat curve whereas heat capacity 
of Ga$_{31}^{+}$ has a well defined peak and has been termed as `magic melter'. 
A similar size sensitive feature has also been observed in the 
case of Al clusters~\cite{Al-expt}. 

The explanation and understanding of various experimental observations have come from the first principles
Density Functional (DF) simulations~\cite{tinthermo,Chacko-PRB,Ga-theo,sailaja-Ga,snsi}.  
For example, the higher than bulk melting temperature 
for Sn and Ga clusters is understood as due to the difference in the nature of bonding 
between the cluster and the bulk~\cite{tinthermo,Ga-theo,snsi}.
However, the extreme size sensitivity displayed in Gallium and Aluminum clusters is still 
an unexplained phenomena. 
The present work addresses this issue by employing first principles  DF methods. 
In this letter we report our results of {\it ab initio} molecular dynamical (MD) simulations 
carried out on Ga$_{30}$ and Ga$_{31}$.
It is of some interest to note that similar size sensitive heat capacities have been observed 
in case of Ga$_{n}$ ($n=17,20$)~\cite{sailaja-Ga} and Na$_{n}$ ($n=40,50,55$)~\cite{Lee-JCP} clusters.
In both these cases addition of  few atoms 
changes the nature of heat capacities significantly.
By analyzing the geometry of the ground state, we establish a definitive correlation between the  nature of 
the ground state and  the observed heat capacity.
Our detailed calculations show that an `ordered' ground state leads to a heat capacity
with a well defined peak while a cluster with `disordered' ground state leads to a 
flat heat capacity with no distinct melting transition.
In what follows we will make the  meaning of `order' and `disorder' precise and provide explanation 
for the size sensitive heat capacities.

We have carried out isokinetic Born-Oppenheimer MD simulations using
ultrasoft pseudopotentials within the generalized gradient approximation (GGA)~\cite{vasp}.  
For all the clusters reported here we have obtained at least 200 equilibrium structures.
For computing heat capacities of Ga$_{30}$ and Ga$_{31}$ the MD calculations were carried out for
16 different temperatures, each with the duration of 150~ps or more, in the range of $100
\le T \le 1100$~K, which results in a total simulation time of 2.4~ns.
In order to get converged heat capacity curves especially in the region of 
coexistence, more temperatures were required with longer simulation times.
The resulting trajectory data has been used to compute the ionic specific heat 
by employing the Multiple Histogram (MH) method~\cite{MH,amv-review}.  

\begin{figure}
\epsfxsize=0.5\textwidth
\centerline{\epsfbox{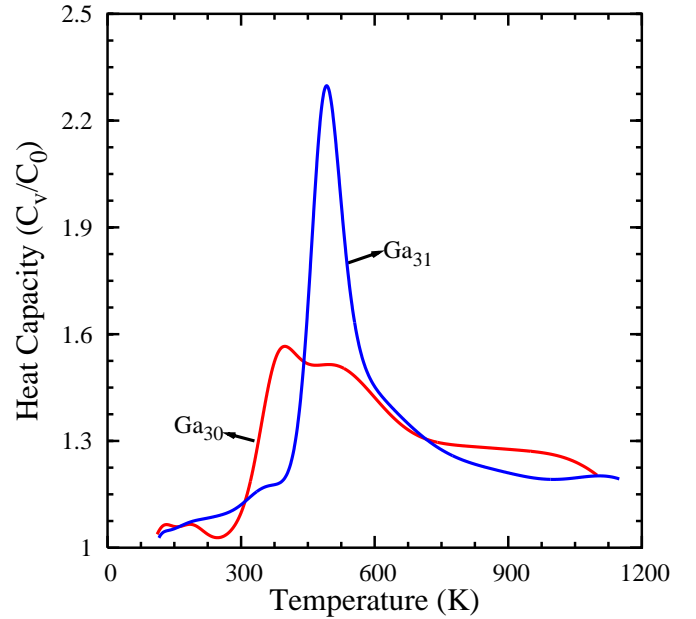}}
\caption{\label{fig1} The heat capacity of Ga$_{31}$, Ga$_{30}$ computed over 90~ps.}
\end{figure}
Fig.\ \ref{fig1} shows the calculated heat capacity of Ga$_{30}$ and Ga$_{31}$.
Evidently the dramatic difference in the heat capacities of Ga$_{30}$ and Ga$_{31}$, observed in 
the experiments is well reproduced in our simulations.
Thus Ga$_{31}$ has a well defined  peak in the heat capacity whereas the heat capacity for Ga$_{30}$ 
is rather flat.
We also note that both Ga$_{30}$ and Ga$_{31}$ becomes `liquid-like' at temperatures much higher
than Tm$_{\rm [bulk]}$ (303~K) {\it i.e.} around 500~K, consistent with the experiments. 
\begin{figure}
\epsfxsize=0.45\textwidth
\centerline{\epsfbox{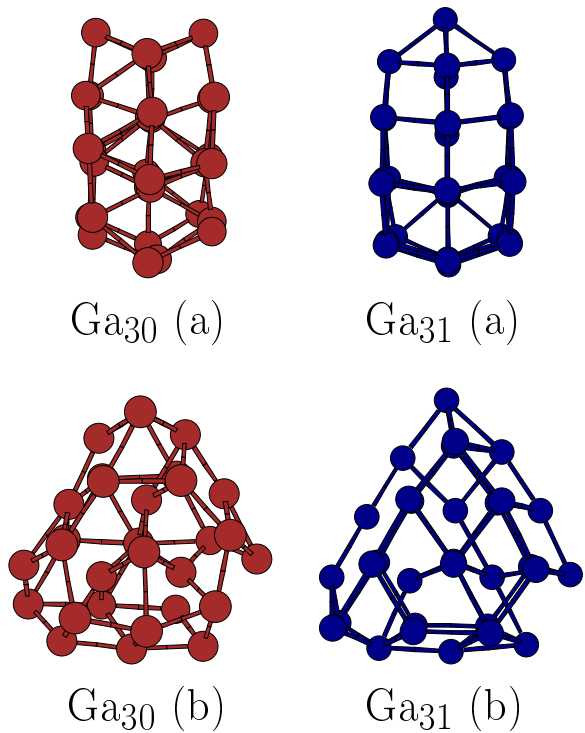}}
\caption{\label{fig2} The ground  state geometry of Ga$_{30}$ and  Ga$_{31}$ with two different perspectives. 
perspective (b) is rotated by 90$^0$ with respect to perspective (a). }
\end{figure}
In order to gain insight in to these observations we
analyze the ground state of Ga$_{30}$ and Ga$_{31}$.
In Fig. \ref{fig2} we show the ground state geometries of Ga$_{30}$ and Ga$_{31}$ 
with two different perspectives.
A cursory  analysis of Fig.\ \ref{fig2}-a may lead to the conclusion that 
the only difference between Ga$_{30}$ and Ga$_{31}$ 
ground states is the presence of the capped atom in Ga$_{31}$.
However, a  different view obtained by rotating the cluster by 90$^0$ brings out the 
significant differences in Ga$_{30}$ and Ga$_{31}$, clearly indicating that  Ga$_{31}$ is more ordered. 
A careful examination of Fig.\ \ref{fig2}-b shows the presence of well ordered planes in Ga$_{31}$. 
Such planes are only in a formative stage and considerably deformed in Ga$_{30}$.
In fact an addition of just one atom in Ga$_{30}$ displaces {\it all} the atoms by a significant
amount which makes Ga$_{31}$ more ordered. 
That a single atom makes a substantial change is also seen by the fact that there is 
noticeable difference in the coordination number in these two clusters. In Ga$_{30}$, 
5 atoms have 4 or more coordination number whereas in Ga$_{31}$, 14 atoms have 4 fold or more 
coordination. Therefore we termed Ga$_{30}$ as a `disordered' structure relative to Ga$_{31}$.

Thus when the system is disordered or  amorphous  each atom (possibly a group of atoms) 
is  likely to have different local environment.
That means different atoms are bonded with the rest of the system with varying strength.
Consequently, their dynamical behavior as a response to temperature will differ.
Some of the atoms may pickup kinetic energy at low temperatures while the others may do so at higher 
temperatures.
In a given structure if a large group of atoms are bonded together with a similar 
strength forming an island of local order it is reasonable to expect that they will 
`melt'  together. In this case the cluster can be considered as (at least partially) ordered 
and will show a well defined peak in the heat capacity. However, if the system is disordered in the sense 
that there are no such islands of significant sizes having  local order then we expect
a very broad continuous phase transformation. Indeed, our analysis of 
mean square displacements (MSDs) for individual atoms
brings out this fact clearly. The MSDs for individual atoms is defined as
\begin{equation}
\langle {\bf r}_{I}^{2}(t)\rangle =\frac{1}{M}
\sum_{m=1}^{M}
\left[ {\bf R}_{I}(t_{0m}+t)-{\bf R}_{I}(t_{0m})\right]^{2},
\label{eqn:msq}
\end{equation}
where ${\bf R}_{I}$ is the position of the  $I$th atom and 
we average over $M$ different time origins $t_{0m}$ spanning
the entire trajectory.
\begin{figure}
\epsfxsize=0.45\textwidth
\centerline{\epsfbox{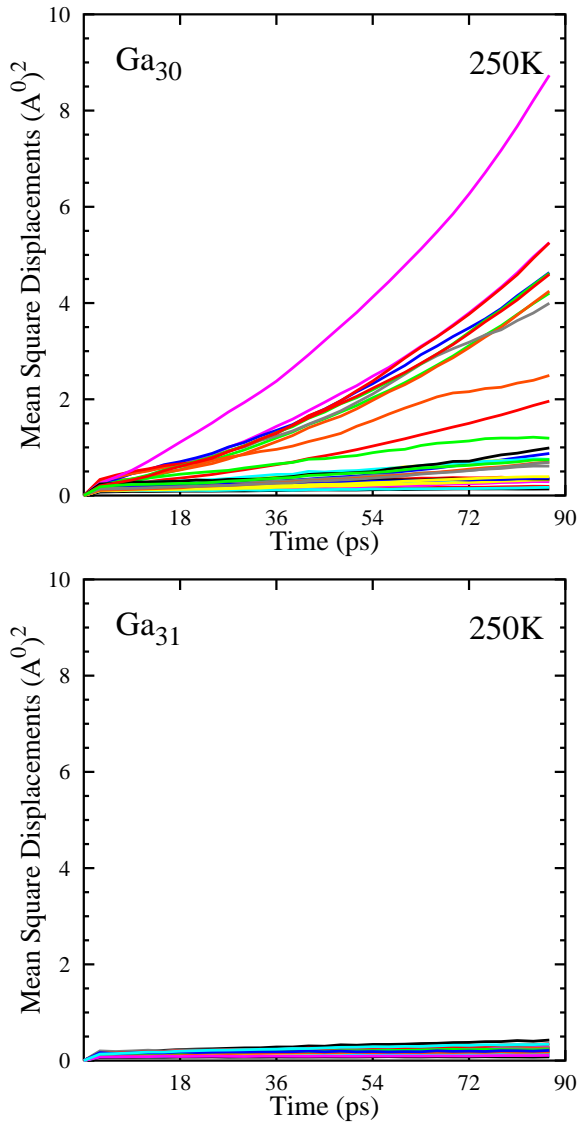}}
\caption{\label{fig3} The MSDs for individual atoms of Ga$_{30}$ and  Ga$_{31}$ computed over 90~ps.}
\end{figure}
In Fig.\ \ref{fig3} we show MSDs of individual atoms for Ga$_{30}$ and Ga$_{31}$
at 250~K.
The contrast between the kinetic response of individual atoms in Ga$_{30}$ and Ga$_{31}$ is very clear. 
For Ga$_{30}$, the MSDs of individual atoms show that some of the atoms (at least 10)
have picked up more kinetic energy compared to others and hence have significantly 
higher displacements (9.0 \AA$^2$ as compared to 0.45 \AA$^2$) whereas in Ga$_{31}$ 
all atoms are oscillating about their mean positions 
and exhibit  small values of MSDs (0.45 \AA$^2$). 
Thus MSDs clearly indicate that in Ga$_{30}$ different atoms have different mobilities.
This wide distribution of MSDs in Ga$_{30}$ indicates that the cluster is in coexistence phase 
around 250~K and is continuously evolving.
This is precisely what is expected if the 
cluster is disordered in the sense described above.
This phenomena has also been observed in the extended systems and
a similar analysis has been used to characterize the nature of spatial inhomogeneities 
with considerable success.~\cite{Chandan-PRL}

\begin{figure}
\epsfxsize=0.5\textwidth
\centerline{\epsfbox{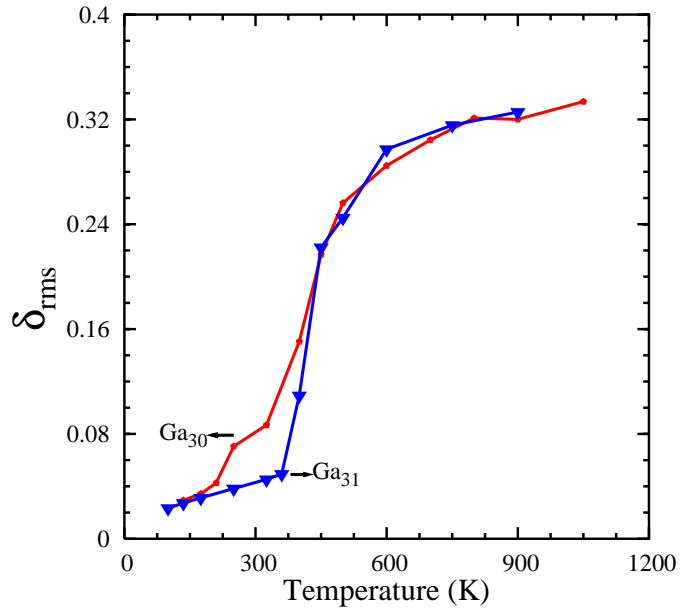}}
\caption{\label{fig3.1} The $\delta_{\rm rms}$ for Ga$_{30}$ and  Ga$_{31}$ computed over last 90~ps.}
\end{figure}
The difference in the mobilities of individual atoms in these two clusters is also reflected 
in the root-mean square bondlength fluctuations ($\delta_{\rm rms}$) shown in Fig.\ \ref{fig3.1}. 
$\delta_{\rm rms}$ shows a clear signal for the beginning of the change of phase around 450~K for Ga$_{31}$.
However, in case of Ga$_{30}$ the transition is spread over a much broader range of temperatures. In fact the 
coexistence region for Ga$_{31}$ is over 175~K (from 425~K to 600~K) and for Ga$_{30}$ it extends over 425~K.
It is interesting to note that in Ga$_{30}$
the isomerization begins around 175~K and continues till 600~K.

The nature of the `order' can also be brought out by examining the Electron Localization Function (ELF).
ELF has been found to be extremely useful for elucidating the bonding
characteristics of a variety of systems~\cite{elf}.  For a single determinantal wavefunction
built from KS orbitals $\psi _{i}$, the ELF is defined as,
\begin{equation}
\chi _{{\rm ELF}}=[1+{(D/D}_{h}{)}^{2}]^{-1},
\end{equation}
where
\begin{eqnarray} D_{h}&=&(3/10){(3{\pi
}^{2})}^{5/3}{\rho }^{5/3}, \\ D&=&(1/2)\sum_{i}{\
|{\bm{\nabla} \psi _{i}}|}^{2}-(1/8){|{\bm{\nabla}
\rho }|}^{2}/\rho, \end{eqnarray} with $\rho \equiv \rho
({\bf r})$ is the valence-electron density.  
ELF is defined such that values of $\chi _{{\rm ELF}}$ approaching unity 
indicate a strong localization of 
the valence electrons and covalent bonding.
ELF partitions the molecular space into
regions or basins of localized electron pairs or attractors.
At very low values of the ELF all the basins are connected.
In other words, there is a single basin containing all the atoms. As the value
of the ELF is increased, the basins begin to split and finally
we will have as many basins as the number of atoms.
The value of the ELF at which the basins split is a measure of the strength of 
interaction between the different atoms.
\begin{figure}
\epsfxsize=0.45\textwidth
\centerline{\epsfbox{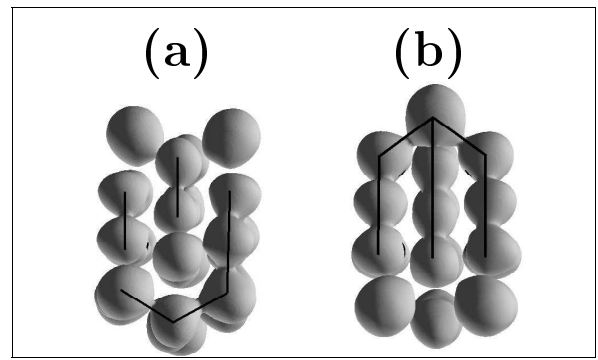}}
\caption{\label{fig4} The Electron Localization Function  of (a) Ga$_{30}$ and  (b) Ga$_{31}$ 
at the value of $\chi_{\rm ELF} = 0.68 $. The black lines show connected basins.}
\end{figure}
Fig.\ \ref{fig4} shows the isosurface of ELF taken at $\chi_{\rm ELF}$ = 0.68, for Ga$_{30}$ 
and Ga$_{31}$. It can be noted that for Ga$_{31}$, 26 atoms are connected via a single basin whereas
for Ga$_{30}$ the largest basin contains 12 atoms with other `fragmented' basins. 
This  supports our earlier observation that Ga$_{31}$ has significantly more similarly bonded
atoms than Ga$_{30}$.
Further evidence for the disordered or amorphous nature of Ga$_{30}$ comes from the 
comparison of entropies of these systems (Fig. not shown). As expected the entropy 
of amorphous structure (Ga$_{30}$) rises rather sharply as compared to Ga$_{31}$ (which is more ordered). 
Quite clearly, the amorphous nature leads to substantially large number  of accessible 
states in case of Ga$_{30}$  and is 
more by a factor of ten as compared to Ga$_{31}$ in the low energy region.

\begin{figure}
\epsfxsize=0.5\textwidth
\centerline{\epsfbox{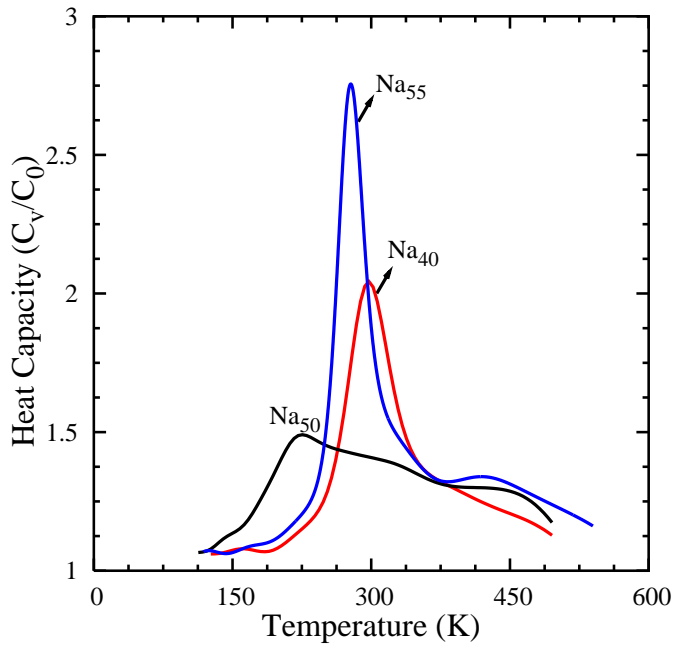}}
\caption{\label{fig6} The heat capacity of Na$_{40}$, Na$_{50}$ and Na$_{55}$ computed over 90ps data.}
\end{figure}

As mentioned earlier this size sensitive behavior is not unique to the 
Gallium clusters reported here and has been observed 
in small Ga~\cite{sailaja-Ga}, Al~\cite{Al-expt}, and Na~\cite{Lee-JCP} clusters.
As an example we show the heat capacities of Na$_n$ ($n= 40,50,55$) clusters in Fig.\ \ref{fig6}. 
The change in the nature of heat 
capacity as the cluster grows from Na$_{40}$ to Na$_{55}$ is quiet evident from Fig.\ \ref{fig6}.
Our detailed analysis of the ground state geometries shows a direct correlation between the 
nature of the ground state and calculated heat capacities. 
It may be noted that Na$_{55}$ is highly symmetric 
and very well ordered. Na$_{40}$ is also ordered and has basin containing substantially large number of atoms
but Na$_{50}$ is relatively disordered which is clearly reflected in their heat capacities~\cite{footnote1}. 

The main contribution of the present work is to bring out a definitive relationship between the local order
in the cluster and its finite temperature behavior. As the cluster grows in size it is very likely 
that it will evolve through a succession of such ordered and disordered geometries. In such
cases addition of one or few atoms is likely to change (as demonstrated in this work) the nature of 
the ground state abruptly. Thus, the size sensitive nature of heat capacities is generic to small 
clusters and related to the evolutionary pattern seen in their ground states. The evidence for this 
comes not only from Gallium clusters but also from clusters of Sodium and Aluminum having 
very different nature of bonding.

Partial support from IFCPAR-CEFPRIA (Project No. 3104-2)  New Delhi and super computing facility from C-DAC 
is gratefully acknowledged.


\begin{thebibliography}{}
\bibitem{footnote1}   It may be noted that these are predicted heat capacities and 
                      await experimental confirmation.
\end{thebibliography}
\end{document}